# Fast multicolor photodetectors based on graphene-contacted p-GaSe/n-InSe van der Waals heterostructures

Faguang Yan[1], Lixia Zhao[2,3], Amalia Patanè[4], PingAn Hu[5], Xia Wei[1], Wengang Luo[1], Dong Zhang[1], Quanshan Lv[1], Qi Feng[1], Chao Shen[1,3], Kai Chang[1,3], Laurence Eaves[4] & Kaiyou Wang[1,3*]

1. *State Key Laboratory of Superlattices and Microstructures, Institute of Semiconductors, Chinese Academy of Sciences, Beijing 100083, China*

2. *State Key Laboratory of Solid-State Lighting, Institute of Semiconductors, Chinese Academy of Sciences, Beijing 10083, China*

3. *College of Materials Science and Opto-Electronic Technology, University of Chinese Academy of Science, Beijing 100049, China*

4. *School of Physics and Astronomy, University of Nottingham, Nottingham NG7 2RD, UK.*

5. *Key Lab of Microsystem and Microstructure, Harbin Institute of Technology, Ministry of Education, Harbin, 150080, China*

∗ Correspondence and requests for materials should be addressed to K.W. (kywang@semi.ac.cn)

**The integration of different two-dimensional materials within a multilayer van der Waals (vdW) heterostructure offers a promising technology for realizing high performance opto-electronic devices such as photodetectors and light sources[1-3]. Transition metal dichalcogenides, *e.g.* $MoS_2$ and $WSe_2$, have been employed as the optically-active layer in recently developed heterojunctions. However, $MoS_2$ and $WSe_2$ become direct band gap semiconductors only in mono- or bilayer form[4,5]. In contrast, the metal monochalcogenides**

**InSe and GaSe retain a direct bandgap over a wide range of layer thicknesses from bulk crystals down to exfoliated flakes only a few atomic monolayers thick[6,7]. Here we report on vdW heterojunction diodes based on InSe and GaSe: the type II band alignment between the two materials and their distinctive spectral response, combined with the low electrical resistance of transparent graphene electrodes, enable effective separation and extraction of photoexcited carriers from the heterostructure even when no external voltage is applied. Our devices are fast (< 10 μs), self-driven photodetectors with multicolor photoresponse ranging from the ultraviolet to the near-infrared and have the potential to accelerate the exploitation of two-dimensional vdW crystals by creating new routes to miniaturized optoelectronics beyond present technologies.**

Multicolor photodetectors covering the ultraviolet (UV), visible and infrared (IR) spectral ranges have potential for a wide range of applications, such as optical communication[8], imaging[9], environmental monitoring[10] and astronomical observations[11]. Furthermore, robust and miniaturized self-driven devices, which require no electrical power source, are of particular interest for applications in extreme conditions. Self-driven multicolor photodetectors using semiconductor heterojunctions, such as $MoS_2/Si$[12] and $MoS_2/GaAs$ heterojunctions[13], have attracted considerable attention recently. Photodetectors based on van der Waals (vdW) heterostructures have also been demonstrated[14,15]. However, self-driven "multicolor" photodetectors are more difficult to realize. VdW heterostructures, which can be assembled by stacking different two-dimensional (2D) semiconductors with different bandgaps, can combine and exploit the properties of the component materials within a single device. Such structures are therefore candidates for multifunctional optoelectronic systems with superior performance. In contrast to gapless graphene[16], GaSe[17] and InSe[18] and monolayers of the transition metal dichalcogenides (TMDCs), are direct band gap semiconductors. High performance photodetectors based on few-layered *p*-type GaSe or *n*-type InSe have been reported previously[19-21]. In addition to their response to visible light, the photoresponse of 2D GaSe photodetectors[17] can extend into the ultraviolet (UV) region, while InSe nanosheets show strong near-infrared (NIR) photoluminescence (PL) emission and photoresponsivity[22]. These results suggest that a heterojunction based on 2D *p*-GaSe and *n*-InSe could be used for photodetection over a still broader spectral range.

Although vdW heterostructures with metal electrodes have been studied extensively and demonstrate interesting optoelectronic and electronic properties[3], their response time ranges from milliseconds[23] to seconds[24]. To fabricate high-performance devices, it is essential that the optically active layers have good interfaces and Ohmic contacts. In contrast to metal-contacted photodetectors, the near perfect optical transparency and the unique electronic properties of graphene make it an ideal electrode for "vertical" vdW heterostructures as it can act as a short, atomically thin charge extraction channel with a large active area, thus enabling both fast and efficient photodetection[21]. An effective method to create faster vdW heterostructure photodetectors is therefore to sandwich the heterojunction between two graphene sheets. These heterostructures can be fabricated with clean interfaces free from dangling bonds, with low defect density and without the Fermi-level pinning that often occurs when metal contacts are directly deposited onto a semiconductor surface.

Here we report on graphene contacted *p*-GaSe/*n*-InSe heterojunctions. A typical device structure is shown schematically in Fig. 1a. A GaSe layer is placed directly on top of the InSe sheet. This sequence of layers ensures that photons of energy $hv < 2$ eV are transmitted through the wide band gap energy GaSe ($E_g = 2.05$ eV at 300K) and can excite electron-hole pairs in the smaller band gap InSe ($E_g = 1.26$ eV at 300K). This large area (~ 50 μm$^2$) *p*-GaSe/*n*-InSe heterojunction exhibits a strong self-driven photoresponse ranging from the UV to NIR due to the built-in potential in the heterojunction, the type-II band alignment between the two layered crystals[7] and their distinctive band gap energies. Furthermore, our use of graphene rather than metals enables a response time as short as 1.85 μs, 3 to 5 orders of magnitude faster than previously reported for photodetectors based on GaSe[17,25] and InSe[21,26].

Figure 1b shows high-resolution transmission electron microscopy images and electron diffraction patterns of the *β*-GaSe and *β*-InSe layers. These have high crystalline quality and in-plane hexagonal symmetry. Their crystal structure consists of Se-M-M-Se (M represents Ga- and In- atoms) layers, as shown in the Supplementary Fig. 1. The measured in-plane lattice constants of GaSe and InSe are $a = 0.37$ and $0.4$ nm, respectively. The separations of two neighboring tetralayers of GaSe and InSe are $d = 0.9$ and $0.84$ nm, respectively. Room temperature measurements of the electrical properties of the *p-n* heterojunction diode reveal strong rectification in the current-voltage (*I-V*) characteristics, with a significant current passing only when the *p*-type GaSe is positively biased (Fig. 1d). The rectification ratio, defined as the ratio of the forward/reverse current, reaches ~10$^5$ at $V_{ds}= +2/-2$ V (Fig. 1d, inset), demonstrating that a good *p-n* diode is formed within the atomically thin GaSe/InSe heterojunction.

Figure 2 shows the dependence of the *I-V* characteristics on light intensity $P$ ranging from 0 to 50 mW cm$^{-2}$. The source-drain current $I_{ds}$ increases with increasing $P$ (Fig. 2a) and the photocurrent $I_{ph}$ exhibits sublinear behavior, *i.e.* $I_{ph} \propto P^{\alpha}$, where α = 0.84, 0.80 and 0.45 at source-drain voltages of $V_{ds}$ = -2, 0 and 2 V, respectively (Fig. 2b). A similar sublinear response has also been reported in other nanomaterials, such as GaN nanowires[27] and WSe$_2$/Graphene heterostructures[28]. Figure 2c,d plot the photoresponsivity ($R=I_{ph}/PS$) and detectivity ($D^*=RS^{1/2}/(2eI_{dark})^{1/2}$) of the heterojunction at different applied voltages as a function of incident light intensity. Here $S$ is the in-plane area (50 μm$^2$) of the device, $e$ is the electron charge and $I_{dark}$ is the dark current. Both $R$ and $D^*$ increase with decreasing light intensity, consistent with the sublinear behavior of the photocurrent.

The response of a photodetector is determined by a combination of several processes, including the excitation, recombination, and diffusion of carriers. Due to the built-in potential of the heterojunction and the type II band alignment (Fig. 3a), at zero bias the photocreated electrons and holes are swept in opposite directions across the junction into the graphene electrodes (Fig. 3b). Thus our devices can operate at zero bias with a photoresponsivity of up to $R$ = 21 mA W$^{-1}$ with corresponding detectivity $D^*$ = 2.2×10$^{12}$ Jones at $\lambda$ = 410 nm. The systematic decrease of $R$ and $D^*$ with increasing $P$ can arise from stronger Coulomb interactions between the photogenerated carriers and enhanced radiative/non-radiative recombination. A reverse bias voltage increases $I_{ph}$ and $R$ due to the increased electric field in the junction, which decreases the carrier transit time, resulting in reduced carrier recombination (Fig. 3c). We also note that $I_{ph}$ and $R$ are strongly enhanced in applied forward biases beyond the open circuit voltage $V_{oc}$ (e.g. $I_{ds} = 0$) and at high $V$. In this regime due to the high injection of majority carriers across the junction, the influence of non-radiative recombination is weaker and a larger number of photogenerated carriers are effectively extracted across the thin layers into the graphene electrodes (Fig. 3d,e).

In our devices, a photoresponsivity of up to $R = 350$ A W$^{-1}$ is obtained at $V_{ds} = 2$ V with an illumination intensity $P = 0.025$ mW cm$^{-2}$ and $\lambda = 410$ nm. This value of $R$ is 2 to 3 orders of magnitude larger than for heterojunction photodetectors based on transition-metal dichalcogenides (TMDCs) such as MoS$_2$/WSe$_2$[3,29] and MoTe$_2$/MoS$_2$[23]. The corresponding detectivity is estimated to be $D^* = 3.7 \times 10^{12}$ Jones, which is two orders of magnitude higher than that of InGaAs/InGaP-based[30] and MoS$_2$-based[31] photodetectors, and is similar to that of Si p-n junction photodetectors[32]. These high $R$ and $D^*$ values indicate that the p-GaSe/n-InSe vertical heterojunction is extremely sensitive to small optical input signals. Furthermore, these devices can operate with no externally applied voltage, thus they have potential for applications that require miniaturized devices with minimal energy consumption.

The spectral response of the p-GaSe/n-InSe heterojunction in Fig. 4a demonstrates a strong photoresponsivity over the range $\lambda = 270$-920 nm, from UV to NIR, under both reverse and zero biases. The photoresponsivity in both reverse bias, $V_{ds} = -2$ V, and zero bias display a similar wavelength dependence. The peak in the photoresponsivity between 400 and 500 nm corresponds to excitations between the $p_{xy}$-like orbitals at the top of the valence band of GaSe and the minimum of the conduction band of InSe. The UV response at $\lambda = 270$ and 350 nm is due to interband optical absorption in the GaSe layer, as for the case of GaSe-based photodetectors[17]. The photoresponse in the NIR wavelength range arises from interband transitions in the InSe layer, which has a band gap of 1.26 eV at room temperature[22]. Based on the photocurrent and incident laser power, we can determine the external quantum efficiency, *EQE*, of the photon to electron conversion (Fig. 4a). The *EQE* ($= hcR/e\lambda$) is defined as the ratio of the number of carriers collected by the electrodes to the number of incident photons, and is wavelength dependent. At zero bias, the device has a maximum *EQE* of 9.3% at $\lambda = 410$ nm, higher than for monolayer $MoS_2$/Si *p-n* diodes.

To further explore the origin of the photoresponse, photocurrent maps were acquired at both zero and reverse biases. Figure 4b shows an optical microscope image of the Gr/GaSe/InSe/Gr heterostructure depicting the relative positions of the GaSe and InSe layers and of the graphene electrodes. The corresponding normalized photocurrent maps at zero and reverse biases with $\lambda = 410$ nm laser excitation (20 µW) are shown in Fig. 4c,d, respectively. To distinguish the different parts of the heterostructure, the GaSe sheet region is outlined in green, the InSe sheet region in purple, and the top and bottom graphene electrodes in solid and dotted gold lines, respectively. The photocurrent map shows that the photosensitive region corresponds to the area where the four component layers (Gr/GaSe/InSe/Gr) are superimposed and demonstrates the formation of a *p-n* junction across the area of the GaSe/InSe interface and the efficient extraction of carriers into the graphene electrodes. The weak photocurrent from the non-overlapping regions shows that the photogenerated carriers in the regions outside the *p-n* junction are separated and extracted less efficiently, even under a reverse bias $V_{ds} = -2$ V (Fig.

4d).

The response time is another important indicator of the performance of a photodetector. To assess their behavior in the time domain, the devices were illuminated with pulsed light generated by a light-emitting diode (LED) ($\lambda$ = 470 nm) powered by a square-wave signal generator. As shown in Fig. 5a, the dynamic response of the photocurrent at $V_{ds}$ = 0 V is described well by the equations $I(t) = I_0[1 - \exp(-t/\tau_r)]$ and $I(t) = I_0\exp(-t/\tau_d)$, where $\tau_r$ = 5.97 µs and $\tau_d$ = 5.66 µs are the rise- and decay-time constants. The even faster photoresponse at $V_{ds}$ = -2 V with $\tau_r$ = 1.85 µs and $\tau_d$ = 2.05 µs (Fig. 5b) is due to the enhanced electric field in reverse bias. Figures 5c,d show that the photocurrent can be switched on and off repeatedly and reproducibly with a square-wave modulation of the light intensity for different laser wavelengths ($\lambda$ = 270, 350, 410, 485, and 570 nm) at a power $P$ = 1 mW cm$^{-2}$. Similar switching behavior is observed for NIR photo-excitation ($\lambda$ = 920 nm) under different illumination intensities for both zero and reverse biases (see Supplementary Fig. 2). ON/OFF ratios up to $10^3$ are observed, demonstrating that the heterojunction can be used as a sensitive and fast photo-detector. The measured response times are significantly faster than those recently reported for van der Waals heterojunction photodetectors[14,21,23,24,33] and Si-based heterojunction photodetectors[34].

In conclusion, we have fabricated and studied Gr/GaSe/InSe/Gr vertical van der Waals heterojunctions with graphene layers acting as transparent electrodes. These devices exhibit rectifying electrical characteristics with a rectification ratio up to ~$10^5$, and self-driven, multicolor photoresponse ranging from the UV to the NIR. The low energy consumption and simple heterostructure design of this nanometer-scale device are particularly notable features. Furthermore, the *p*-GaSe/*n*-InSe photodetector possesses a fast response time down to 1.85 μs and a high photoresponsivity of up to 350 A W$^{-1}$, with corresponding detectivity of $3.7 \times 10^{12}$ Jones. These features point the way to fast, efficient multicolor optoelectronic devices based on multilayer Gr/GaSe/InSe/Gr heterostructures for miniaturized optoelectronic applications that require low energy consumption. Our findings should also stimulate research into the synthesis and fabrication of these materials for future commercial technologies.

**Methods**

**Fabrication of the vertical heterostructure devices.** The few-layer GaSe and InSe flakes were mechanically exfoliated using adhesive tape from bulk single crystals of *β*-GaSe and *β*-InSe. The monolayer graphene, which is used as a transparent electrode, was synthesized by chemical vapour deposition (CVD) on a Cu substrate. The CVD graphene microstamps were first transferred onto a fused silica substrate. The InSe was mechanically exfoliated onto a stamp, which was adhered to a glass slide to facilitate handling and identification by optical microscopy. The target InSe sheet was then transferred to the top of the graphene electrode. Using the same method, the GaSe sheet was transferred on top of the InSe sheet to fabricate a 2D heterojunction. Finally, another CVD graphene microstamp was transferred onto the GaSe sheet to form the top electrode. Metallic contacts were fabricated on the substrate prior to the transfer using standard photoetching, thermal evaporation and lift-off. A fused silica substrate was used to avoid optical interference effects in the SiO$_2$/Si substrate. The active overlap area of this vertical device is typically around 50 μm$^2$. The schematic illustration of the *p*-GaSe/*n*-InSe device is presented in Fig. 1a. The thickness of both GaSe and InSe flakes was determined by atomic force microscopy (Bruker Multimode 8). The typical thickness used for the sensitive photodetectors in this work is 10-20 nm. All the measurements were performed in ambient atmosphere and at room temperature.

**Electrical and opto-electrical studies.** Electrical and opto-electrical measurements were acquired using an Agilent Technology B1500A and a lock-in amplifier (SR830) with a light chopper. The UV and visible to NIR monochromatic illumination were provided by a Zolix Omni-λ300 monochrometer with Xe lamp and a Fianium WhiteLase Supercontinuum Laser Source with repetition rate of 20 MHz, respectively. The output laser wavelength was tuned by an Omni-λ300 monochromator. A microscope objective (Olympus SLMPLN100×) and an micromechanical stage were used to carry out the spatially resolved photocurrent mapping. A home-built measurement system that combined a LED ($\lambda$ = 470 nm), an oscilloscope, and a pulse generator was used to investigate the photoresponse of the device under pulsed light illumination at different frequencies. The response times of the photodetectors were evaluated by a square voltage pulse light and a digital oscilloscope, (Tektronix TDS 2022B).

**Acknowledgments**

This work was supported by '973 Program' No. 2014CB643903, by the NSFC Grant No. 61225021, 11474272, 11174272, and 11404324, by the EU Graphene Flagship Project, and the Engineering and Physical Sciences Research Council (Grant No. EP/M012700/1). The Project was also sponsored by K. C. Wong Education Foundation.

**Author contributions**

K.W. conceived and designed the experiments. F.Y., L.Z., X.W. and Q.L. conducted the measurements. P.H. synthesized the GaSe and InSe crystals. F.Y., W.L., L.Z., A.P. and K.W. wrote the manuscript. All authors discussed the results and revised the manuscript.

**Competing financial interests**

The authors declare no competing financial interests.

**FIGURES**

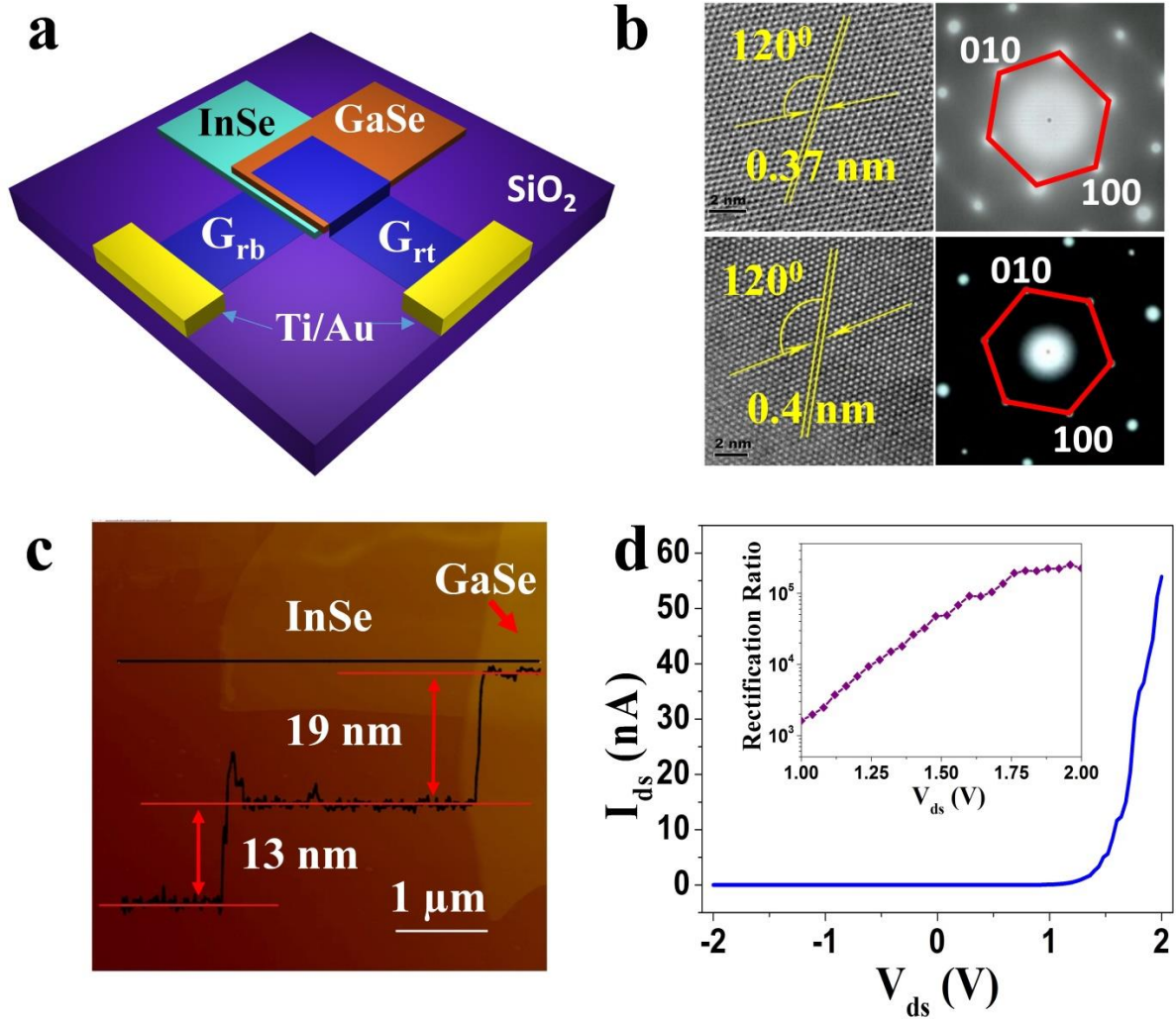

**Figure 1 | Schematic diagram and current-voltage *I-V* characteristic of the *p*-GaSe/*n*-InSe heterojunction diode. a,** Schematic diagram of the *p*-GaSe/*n*-InSe heterojunction diode. **b,** High-resolution TEM image of the GaSe (top left) and InSe (bottom left), respectively. Images on the right show the corresponding electron-beam diffraction patterns of GaSe and InSe. **c,** AFM image of the device. The inset shows the thickness of the different layers. **d,** The *I-V* characteristic of the *p*-GaSe/*n*-InSe heterojunction diode at room temperature. The inset shows the rectification ratio as a function of the source-drain voltage $V_{ds}$.

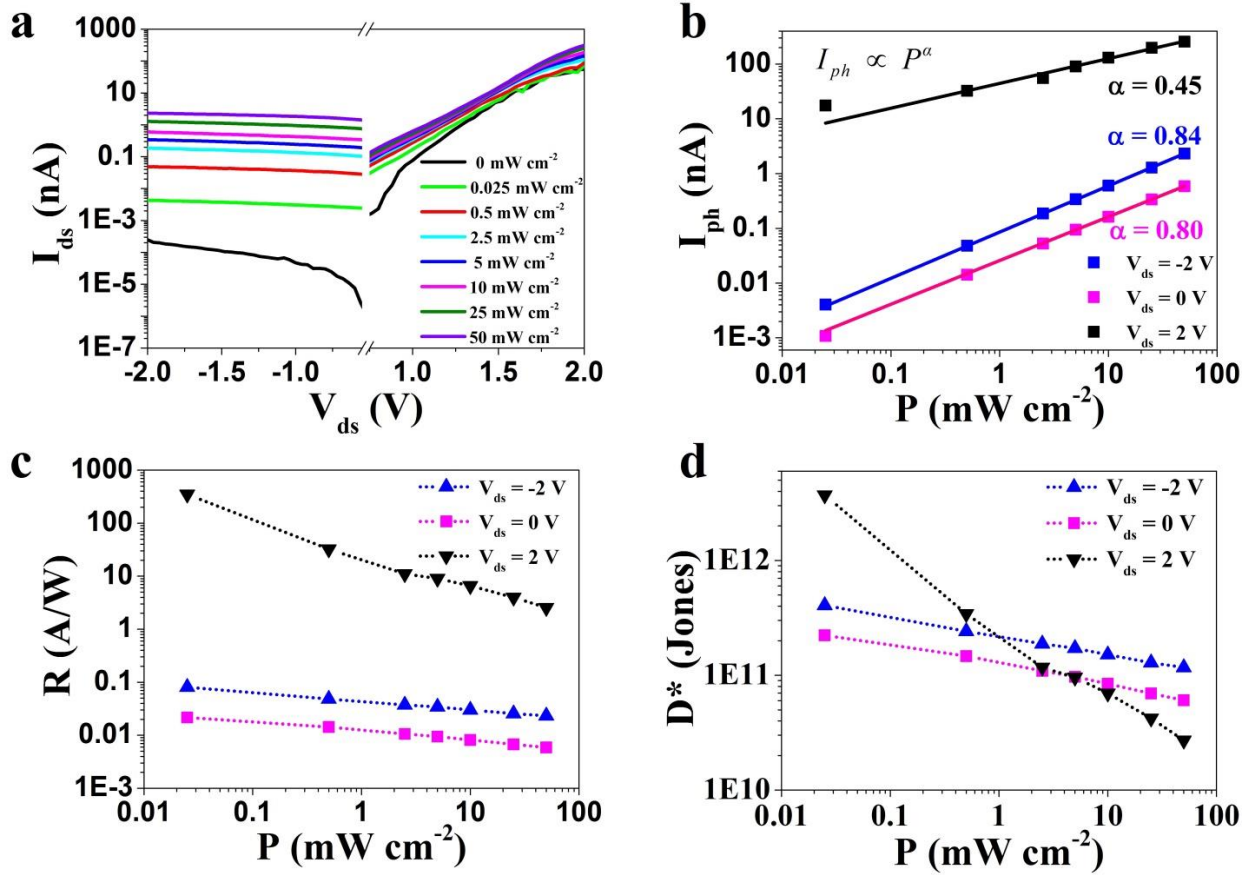

**Figure 2 | Power-dependent optoelectronic characterization at different applied voltages $V_{ds}$. a,** Typical $I_{ds}$ curves of the p-GaSe/n-InSe heterojunction diode with illumination at various excitation intensities ($P$ = 0, 0.025, 0.5, 2.5, 5, 10, 25, 50 mW cm$^{-2}$) and wavelength $\lambda$ = 410 nm at room temperature. **b,** Photocurrent as a function of the illumination intensity at different $V_{ds}$ (forward bias $V_{ds}$ = 2 V, zero bias $V_{ds}$ = 0 V and reverse bias $V_{ds}$ = -2 V). The solid lines are fits to the data. **c,d,** Photoresponsivity $R$ (**c**) and detectivity $D^*$ (**d**) of the heterojunction diode as a function of the illumination intensity $P$ at different $V_{ds}$ ($V_{ds}$ = 2, 0, -2 V).

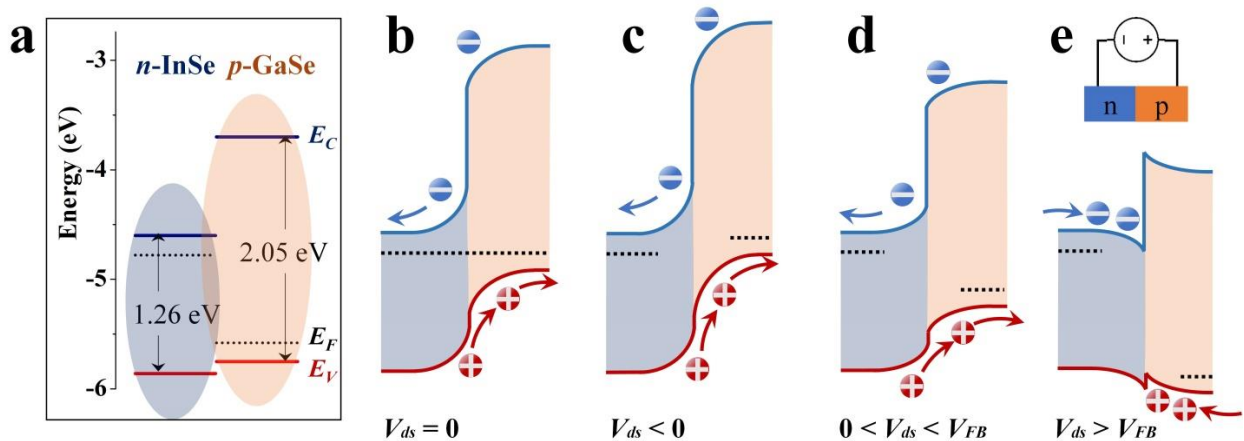

**Figure 3 | Band alignment at the interface of the *p*-GaSe/*n*-InSe heterojunction. a,** Band alignment for isolated *n*-InSe and *p*-GaSe layers. Electron affinities of InSe and GaSe are $\chi =$ -4.6 and -3.7 eV, respectively. The conduction minimum (CB) of GaSe lies above that of InSe by $\Delta E_c = 0.9$ eV whereas the valence band (VB) edge of InSe lies below ($\Delta E_v = -0.1$ eV) that of the larger band gap GaSe, resulting in a type II band alignment. **b,c,d,e,** Schematic band alignment at the interface of the heterojunction at different applied voltages $V_{ds}$ (reverse bias **(c)**, zero bias **(b)** and forward bias **(d,e)**). $V_{FB}$ is the voltage corresponding to the flat band condition.

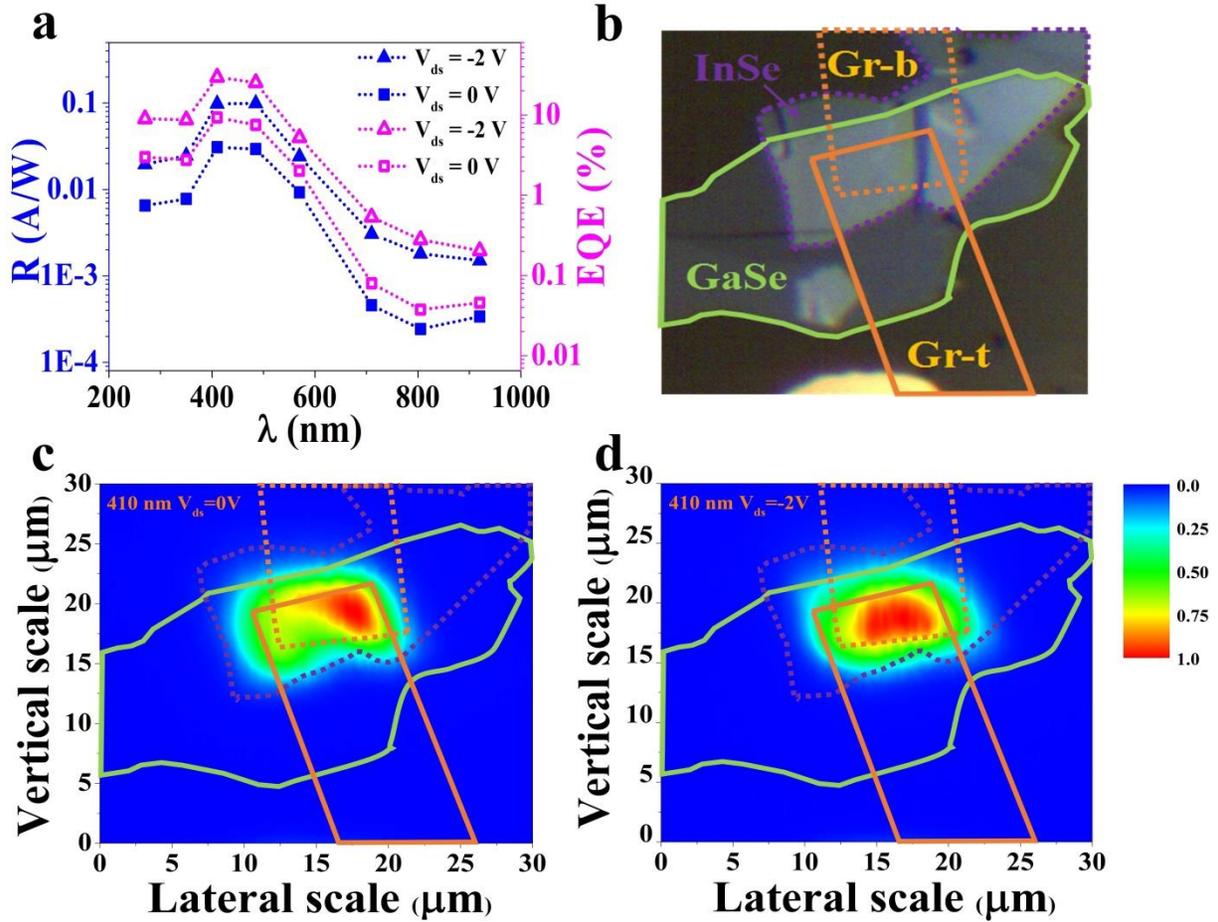

**Figure 4 | Spectral responsivity and normalized photocurrent maps of the *p*-GaSe/*n*-InSe heterojunction. a,** Room temperature photoresponsivity, *R*, and external quantum efficiency, *EQE*, as a function of illumination wavelength at different $V_{ds}$ ($V_{ds}$ = 0 V and $V_{ds}$ = -2 V) and illumination intensity $P$ = 1 mW cm$^{-2}$. **b,** Optical microscope image of the heterojunction; t and b refer to top and bottom graphene electrodes. **c,d,** Normalized photocurrent maps of the Gr/GaSe/InSe/Gr device obtained by scanning a focused laser beam at $V_{ds}$ = 0 V **(c)** and -2 V **(d)** with wavelength $\lambda$ = 410 nm ($P$ = 20 µW). The green solid line outlines the GaSe sheet, the purple dotted line outlines the InSe sheet and the golden solid and dotted lines outline the top and bottom graphene electrodes, respectively. Photocurrent is observed where the 4 layers overlap.

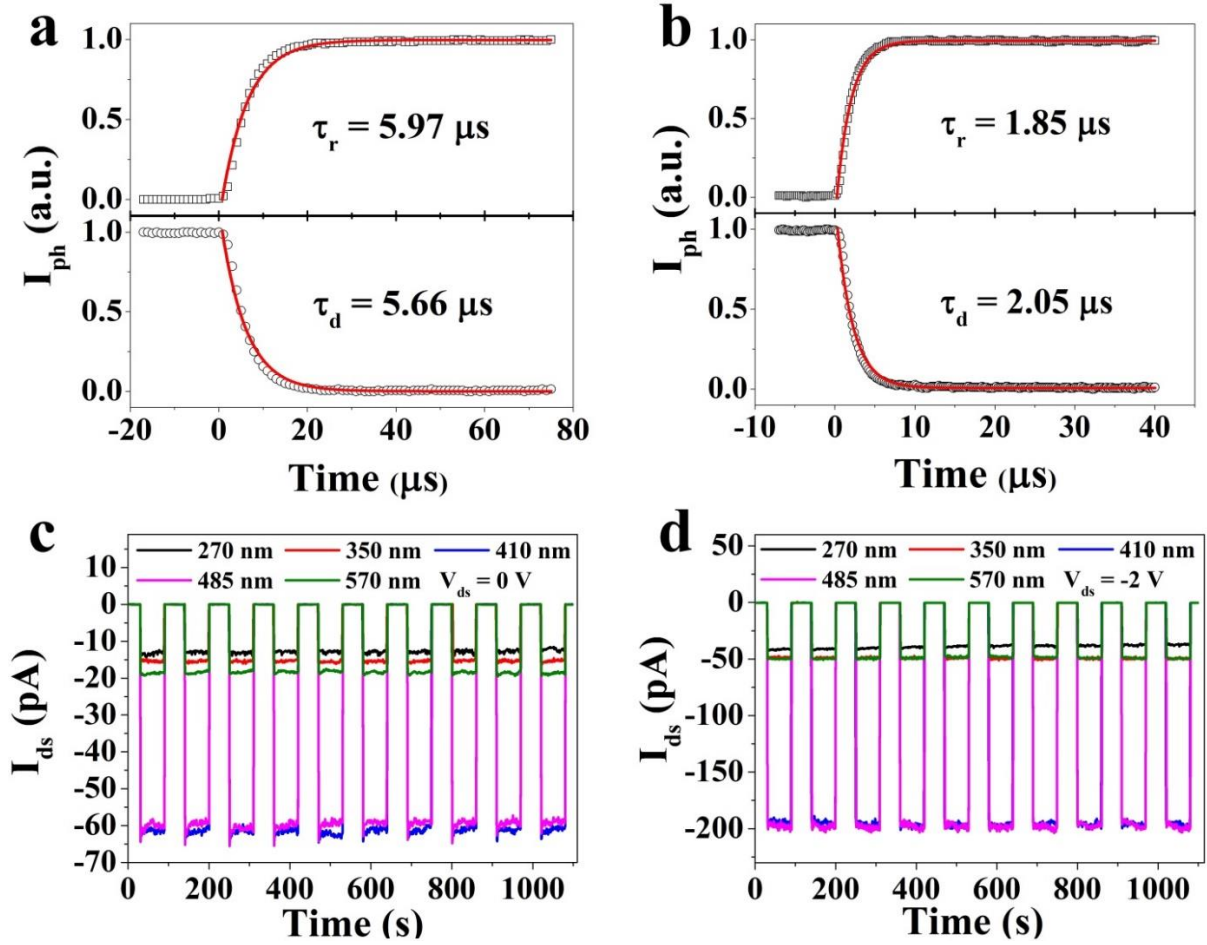

**Figure 5 | Response time and photo-switching of the *p*-GaSe/*n*-InSe heterojunction diode.**
**a,b,** Temporal dependence of the photocurrent and times $\tau_r$ and $\tau_d$ at $V_{ds}$ = 0 V (**a**) and $V_{ds}$ = -2 V (**b**) at room temperature. The red solid lines are fits to the data. **c,d,** Source-drain current $I_{ds}$ as a function of time with photoswitching at $V_{ds}$ = 0 V (**c**) and $V_{ds}$ = -2 V (**d**) under illumination with different wavelengths ($\lambda$ = 270, 350, 410, 485, and 570 nm) and light intensity $P$ = 1 mW cm$^{-2}$.
20

# Supplementary Information

## Fast multicolor photodetectors based on graphene contacted p-GaSe/n-InSe van der Waals heterostructures

Faguang Yan[1], Lixia Zhao[2,3], Amalia Patanè[4], PingAn Hu[5], Xia Wei[1], Wengang Luo[1], Dong Zhang[1], Quanshan Lv[1], Qi Feng[1], Chao Shen[1,3], Kai Chang[1,3], Laurence Eaves[4] & Kaiyou Wang[1,3*]

1. *State Key Laboratory of Superlattices and Microstructures, Institute of Semiconductors, Chinese Academy of Sciences, Beijing 100083, China*

2. *State Key Laboratory of Solid-State Lighting, Institute of Semiconductors, Chinese Academy of Sciences, Beijing 10083, China*

3. *College of Materials Science and Opto-Electronic Technology, University of Chinese Academy of Science, Beijing 100049, China*

4. *School of Physics and Astronomy, University of Nottingham, Nottingham NG7 2RD, UK.*

5. *Key Lab of Microsystem and Microstructure, Harbin Institute of Technology, Ministry of Education, Harbin, 150080, China*

* Correspondence and requests for materials should be addressed to K.W.( kywang@semi.ac.cn)

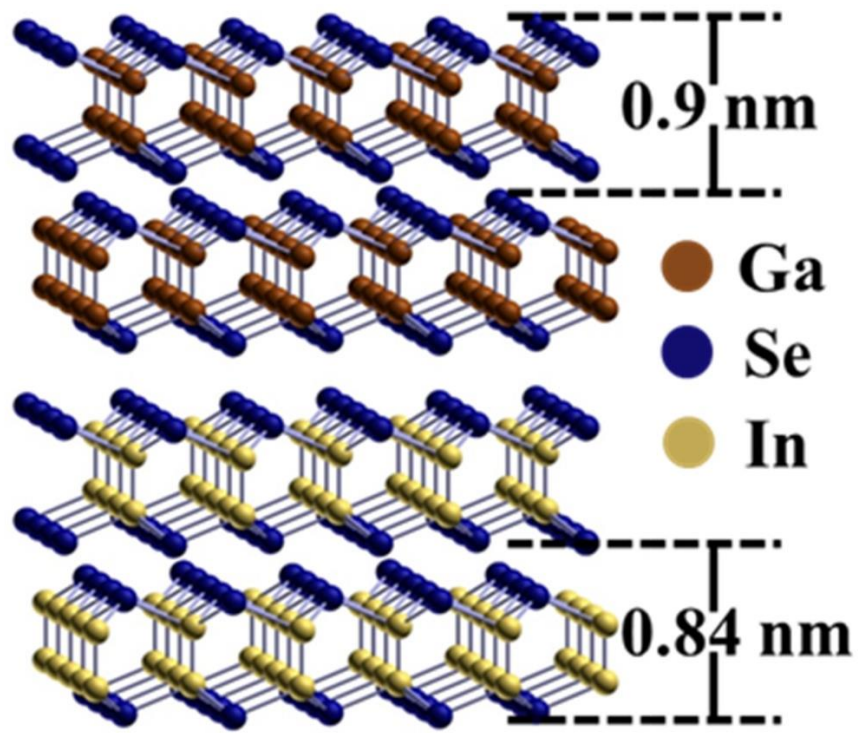

**Supplementary Figure 1 | The structure of GaSe and InSe.** The distances between two neighboring layers of GaSe and InSe are 0.9 and 0.84 nm, respectively.

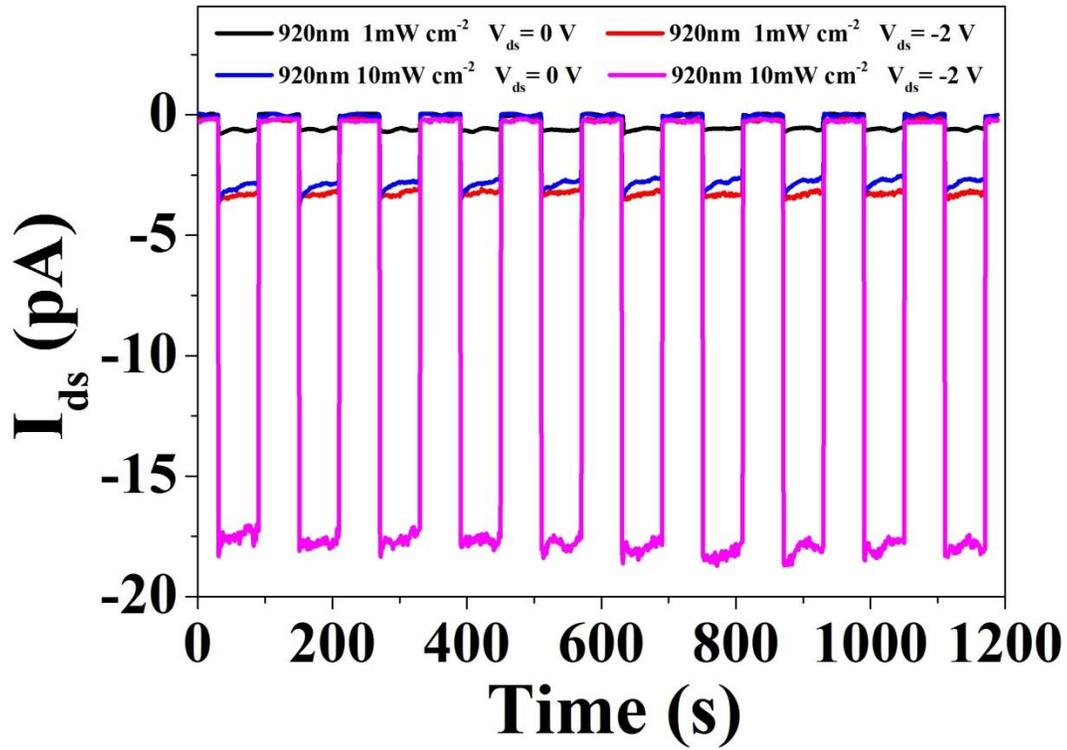

**Supplementary Figure 2 | The photoswitching performance at NIR region.** The $I_{ds}$ of the p-GaSe/n-InSe heterojunction diode as a function of time at $\lambda$ = 920nm under different illuminate intensities for $V_{ds}$ = 0 V and $V_{ds}$ = -2 V. The periodic switching behaviors of the p-GaSe/n-InSe heterojunction diode is obtained by switching the light on and off under global illumination with a 920 nm laser and light intensities of 1 and 10 mW cm$^{-2}$. This demonstrates that the diode can be used as a high quality self-driven photodetector and switcher in the NIR region.